\documentstyle[aps,prl,twocolumn,psfig]{revtex}

\draft
\begin{document}
\newcommand {\be}{\begin{equation}}
\newcommand {\ee}{\end{equation}}
\newcommand {\bea}{\begin{eqnarray}}
\newcommand {\eea}{\end{eqnarray}}
\newcommand {\nn}{\nonumber}

\twocolumn[\hsize\textwidth\columnwidth\hsize\csname@twocolumnfalse%
\endcsname

\title{Theory of c-axis Josephson tunneling in 
$\rm d_{x^2-y^2}$-wave superconductors 
}

\author{Kazumi Maki and Stephan Haas}
\address{Department of Physics and Astronomy, University of Southern
California, Los Angeles, CA 90089-0484
}

\date{\today}
\maketitle

\begin{abstract}
The temperature and angular dependence of the c-axis Josephson current 
and the superfluid density in layered $\rm d_{x^2-y^2}$-wave superconductors
are studied within the framework of an extended
Ambegaokar-Baratoff formalism. In particular, the effects of 
angle-dependent tunneling matrix elements and Andreev scattering at grain 
boundaries are taken into account. These lead to strong corrections of
the low-temperature behavior of the plasma frequency and the Josephson 
current. Recent c-axis measurements on the 
cuprate high-temperature superconductors   
$\rm HgBa_2CaCu_{1+\delta}$ and $\rm Bi_2Sr_2CaCu_2O_{8+\delta}$ 
can therefore be interpreted to be
consistent with a $\rm d_{x^2-y^2}$-wave order parameter.

\end{abstract}
]

Interfaces such as grain boundaries in the high-temperature (high-$\rm T_c$)
superconductors are known 
to have important theoretical, technical, and practical implications.
\cite{hilgenkamp}
In a set of early experiments it was shown that the critical
current between single-crystal surfaces depends sensitively on their relative
orientation.\cite{dimos,ivanov}
More precisely, when two such samples are matched along their crystal 
a-b planes, but twisted with respect to each other about the direction 
of their common c-axis, it was found that the critical current decreases
rapidly as a function of the twist angle $\alpha$. This strong angular 
dependence has since been observed frequently at grain boundaries in the 
high-$\rm T_c$ 
cuprates. It is a consequence of the $\rm d_{x^2 - y^2}$-wave symmetry 
of the superconducting order parameter in these compounds.\cite{hilgenkamp} 

\begin{figure}[h]
\centerline{\psfig{figure=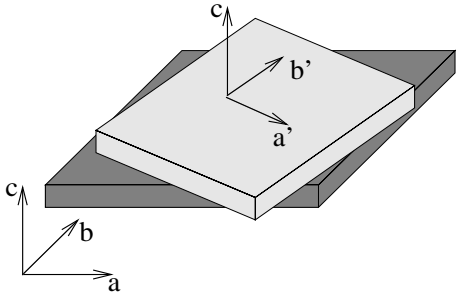,width=6cm,angle=0}}
\vspace{0.3cm}
\caption{
Schematic representation of two single-crystal samples, matched along their a-b
planes and twisted about the c-axis by an angle $\alpha$. 
}
\end{figure}

In the recent literature, there have been theoretical as well as experimental
puzzles regarding the precise nature of c-axis
Josephson tunneling in the cuprate superconductors, particularly in 
$\rm Bi_2Sr_2CaCu_2O_{8+\delta}$.
On the theoretical front, several scenarios have been considered to model
the angular behavior of the Josephson current.\cite{bruder,barash,golubov}
However, in these frameworks it appears difficult to reproduce the strong
dependence on the twist angle $\alpha$ which has been observed in 
experiment. On the 
experimental side, there is a recent controversy arising from a c-axis
Josephson tunneling experiment in which two single crystals were twisted 
about the c-axis and subsequently joined together.\cite{qiang} 
In contrast to earlier experiments\cite{dimos,ivanov}, the authors found a
critical current that is very small but independent of $\alpha$, implying
that the superconducting order parameter in $\rm Bi_2Sr_2CaCu_2O_{8+\delta}$
has a dominant s-wave component.\cite{qiang} However, it appears that the 
tunneling area in this particular
measurement may have been too large to guarantee
homogeneity of the contact areas. Furthermore, the small magnitude of the 
observed Josephson current suggests a large contribution from the 
surface area, which in turn would not adequately reflect the symmetry of
the bulk superconducting order parameter.\cite{tsuei}

We will analyze a set of more recent experiments, 
which measure the c-axis Josephson current in superconducting 
$\rm Bi_2Sr_2CaCu_2O_{8+\delta}$.\cite{takano,takano2} Two single-crystal
whiskers of this compound with very small widths $\Phi \sim 10 \mu m$
were joined along their a-b planes. By varying their relative orientations
the angular dependence of the Josephson current along the c-axis was 
detected. A clear fourfold symmetry with respect to the twist angle
$\alpha$ was observed, consistent with a $\rm d_{x^2 - y^2}$-wave
superconducting order parameter for this compound. Surprisingly, however, 
the angular dependence turned out to be much more pronounced than one would
have expected from previous theoretical treatments, based on the 
Ginzburg-Landau model and on the Ambegaokar-Baratoff formalism for 
superconductor-insulator-superconductor (S-I-S) junctions, which predict 
a simple 
$\cos{(2 \alpha)}$ dependence for $\rm d_{x^2 - y^2}$-wave superconductors.  
\cite{ambegaokar}

Here we propose an extension of the Ambegaokar-Baratoff
treatment for the superfluid density and the Josephson current along the 
c-axis of layered unconventional superconductors, taking into 
account (i) the coherent component of the tunneling current, (ii) the  
$\rm d_{x^2 - y^2}$-wave symmetry of the order parameter, and (iii) the 
angular dependence of the transfer matrix element along the crystal 
c-direction.   

Let us first discuss the issue of coherence. In contrast to the case
of s-wave superconductivity, Josephson coupling between two unconventional 
superconductors with nodes in their gap functions is possible only if 
there is a coherent component to the tunneling current. Hence, purely 
incoherent tunneling will not result in a Josephson coupling between two 
$\rm d_{x^2 - y^2}$-wave superconductors, as already noted in Refs.
\cite{bruder} and \cite{graf}, and for the layered $\kappa$-(ET)$_2$ salts 
in Refs.~\cite{pinteric} and \cite{dora}. Coherent tunneling between adjacent 
layers with planar quasiparticle wave vectors $\vec{k}$ and $\vec{k'}$ can be
defined by the condition $\vec{k} \parallel \vec{k'}$. This is what 
Ambegaokar and Baratoff called ``specular transmission".
\cite{ambegaokar,hirschfeld} On the other hand, if the tunneling is 
incoherent $\vec{k} \neq \vec{k'}$, i.e. $\vec{k}$ and $\vec{k'}$ are 
completely uncorrelated. For the sake of the following 
argument, let us consider only 
these two extreme cases. In real materials, coherent and 
incoherent processes
are typically mixed, in which case only the coherent component is relevant.
The Josephson coupling is proportional to $\langle \Delta (\phi) 
\Delta (\phi ') \rangle$, where $\langle \cdot \cdot \cdot \rangle$ 
represents angular averages over $\phi$ and $\phi '$, which are the 
angles between the wave vectors $\vec{k}$ and $\vec{k}'$ and the 
crystal a-axis. Hence, when $\vec{k}$ and $\vec{k}'$ are uncorrelated 
$\langle \Delta (\phi) \Delta (\phi ') \rangle = \langle \Delta (\phi) \rangle
 \langle \Delta (\phi ') \rangle = 0$. This argument applies not only to 
$\rm d_{x^2 - y^2}$-wave systems, but also to other unconventional 
superconductors. In contrast, there will be Josephson coupling between 
two s-wave superconductors even when the tunneling is completely 
incoherent, and therefore this is one of the important differences 
between conventional and unconventional superconductivity.  

When the transport along the c-axis is due to coherent tunneling, 
the temperature
dependence of the out-of-plane superfluid density $\rho_{s c}$
is different from that of the in-plane  superfluid density $\rho_{s 
\parallel}$. Recent calculations performed for anisotropic 
three-dimensional models have predicted
a linear decrease with increasing temperature, both for $\rho_{s c}$
and $\rho_{s \parallel}$.\cite{won,radtke} However, within the tunneling 
model, the out-of-plane superfluid density for $\rm d_{x^2-y^2}$-wave
superconductors is given by
\bea
\rho_{s c} (t) & = & \frac{\Delta (t)}{\Delta_0} \int_0^{\pi /2}
d\phi \sin{\phi }  \tanh{\left( \frac{\Delta (t) \sin{\phi }}{ 2 T} \right)}  \\
 & \approx  & 1 - 1.645 \left( \frac{T}{\Delta_0}\right)^2 - 
 3.606 \left( \frac{T}{\Delta_0}\right)^3, \nonumber
\eea
where $t=T/T_c$ is the reduced temperature and $\Delta (t) $ is the amplitude
of the superconducting gap function. Thus the tunneling model predicts
a $t^2$-dependence at low temperatures, consistent with the out-of-plane 
superfluid density observed in $\rm YBa_2Cu_3O_{7-\delta}$\cite{hosseini},
$\rm Tl_2Ba_2CuO_6$\cite{tsvetkov}, and $\rm \kappa-(ET)_2Cu(N(CN)_2)Br$
\cite{pinteric}.  

More recently, the out-of-plane superfluid density in 
$\rm Bi_2Sr_2CaCu_2O_{8+\delta}$ was measured, using a sensitive 
microwave resonance technique.\cite{gaifullin}
Noting  that the square of the Josephson plasma frequency is proportional 
to $\rho_{s c} (t)$, it was observed that the low-temperature behavior 
of the c-axis superfluid density is proportional to $t^5 \sim t^6$. A 
comparable high power-law has also  been reported in 
$\rm HgBa_2CaCu_{1+\delta}$.\cite{panagopoulos} In order to describe
this behavior, Xiang and Wheatley\cite{xiang} introduced an 
angle-dependent tunneling matrix element proportional to $\cos^2{(2 \phi)}$,
which vanishes along the nodal directions. This behavior
is common to those high-$\rm T_c$ cuprates which do not contain Cu-O chains. 
It is caused by the overlap of Cu 4s and O 2p orbitals.
\cite{gaifullin,xiang} Introducing an angle-dependent tunneling matrix 
element into the formalism for the superfluid density leads to
\bea
\rho_{s c} (t) & = & \frac{15 \Delta (t)}{8 \Delta_0} \int_0^{\pi /2}
d\phi \sin^5{\phi }  \tanh{\left( \frac{\Delta (t) \sin{\phi }}{ 2 T} \right)}  \\
 & \approx  & 1 - 3.606 \left( \frac{T}{\Delta_0}\right)^3 - 
 443.35 \left( \frac{T}{\Delta_0}\right)^6. \nonumber
\eea
While the leading power appears to be $t^3$, comparison of  
the coefficients of the $t^3$- and $t^6$-terms shows the $t^6$-power
to be dominant at all experimentally relevant low temperatures. 

\begin{figure}[h]
\vspace{0.6cm}
\centerline{\psfig{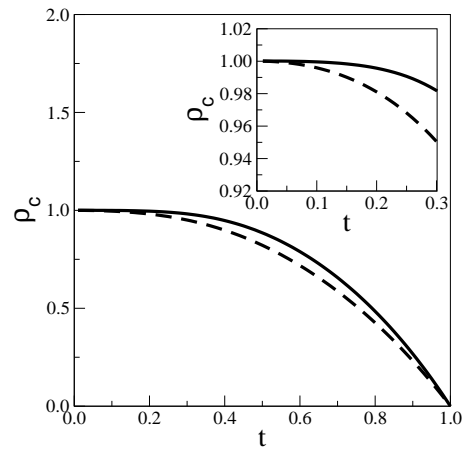}}
\vspace{0.3cm}
\caption{Out-of-plane superfluid density as a function of temperature.
The solid line includes the angular dependence of the tunneling matrix element
along the c-direction, whereas the dashed line does not. The inset shows 
the low-temperature region.
}
\end{figure}

In Fig.~2 the out-of-plane superfluid density of a 
$\rm d_{x^2-y^2}$-wave superconductor is shown as a function of the 
reduced temperature. Comparing the behavior of $\rho_{s c} (t)$ with 
and without inclusion of the angular dependence in the tunneling matrix
element, the difference appears minor to the eye. However, by zooming into the 
low-temperature regime (inset of Fig.~2) one observes that the extra 
$\cos^2{(2 \phi)}$-term leads to a strongly reduced temperature dependence 
in this regime. A similar difference in the low-temperature behavior
has been reported in measurements of the plasma resonance of 
various high-$\rm T_c$ cuprate compounds (Fig.~3 of Ref. \cite{gaifullin}).   

Let us now consider the angular and temperature dependence of the Josephson 
current, as probed using  
$\rm Bi_2Sr_2CaCu_2O_{8+\delta}$ cross-whiskers twisted about their
common c-axis.\cite{takano2}  
By applying the Ambegaokar-Baratoff formalism, and assuming that there is a
component of coherent tunneling along the c-direction, one obtains
\bea
I_c(T,\alpha) & = & \frac{\Delta^2(t)}{\Delta_0} \int_0^{2 \pi} d\phi
\sum_{n=0}^{\infty} \frac{T f_> f_<}{(1 + (\Delta_0 / t_{\perp})^2
(|f_>| - |f_<|)^2)} \nn \\
& \times & \frac{1}{\sqrt{\omega_n^2 + \Delta^2_0
|f_>|^2} \sqrt{\omega_n^2 + \Delta^2_0 |f_<|^2} },
\eea
where $\omega_n = 2\pi T (n + 1/2)$ is the Matsubara frequency, and 
$t_{\perp}$ is the tunneling matrix element along the c-direction. 
The angular dependence of the gap function is contained 
in the factors
\bea
& f_>& = \cos{(2\phi)},\ \ \  f_<=\cos{(2(\phi + \alpha))},\\ 
& \rm if &\ \ 
| \cos{(2\phi)}| > |\cos{(2(\phi + \alpha))}|,\nn \\
& f_>& = \cos{(2(\phi+\alpha))},\ \ \  f_<=\cos{(2\phi )}, \\   
& \rm if & \ \ 
| \cos{(2\phi)}| < |\cos{(2(\phi + \alpha))}| \nn. 
\eea
The factor in the denominator $(1 + (\Delta_0 / t_{\perp})^2
(|f_>| - |f_<|)^2)$ includes the effect of Andreev reflections 
\cite{blonder} at the grain boundary. If this factor is neglected,
one obtains the usual Ambegaokar-Baratoff expression, with 
a dependence of the Josephson current on the twist angle very 
close to $\cos{(2 \alpha)}$. Furthermore, one notes that this correction 
of the Josephson current due to Andreev scattering vanishes if 
$\alpha$ = 0, i.e. when the two whiskers are perfectly aligned.

Taking into account the additional angular dependence of the tunneling matrix 
element in the above derivation leads to a slightly modified expression, 
analogous to the discussion of the superfluid density,
\bea
I_c(T,\alpha) & = & \frac{\Delta^2(t)}{\Delta_0} \int_0^{2 \pi} d\phi
\sum_{n=0}^{\infty} \frac{T f_> f_<}
{(1 + (\Delta_0 / t_{\perp})^2
(|f_>| - |f_<|)^2)} \nn \\
& \times & \frac{15 |f_> f_< |^2}{8 \sqrt{\omega_n^2 + \Delta^2_0
|f_>|^2} \sqrt{\omega_n^2 + \Delta^2_0 |f_<|^2} }.
\eea

Let us first discuss the angular dependence of the zero-temperature 
limit of Eqs.~3 and 6, given by
\bea
I_c (T= 0 , \alpha ) & = & \frac{1}{2\pi} \int_0^{2 \pi} \frac{f_> f_< }{|f_>|}
K \left( \sqrt{1 - \left(\frac{f_<}{f_>}\right)^2}\right)\nn \\
& \times & \frac{F(\phi)}{(1 + (\Delta_0 / t_{\perp})^2 (|f_>| - |f_<|)^2)},
\eea
where $K(k)$ is the complete elliptic integral of the 
first kind.  
Here $F(\phi ) = 1$ for the case of angle-independent tunneling along the 
c-axis
(Eq.~3), and $F(\phi ) = |f_< f_>|^2$ when this angular dependence is included
(Eq.~6). 

Fig.~3 shows the angular dependence of the c-axis Josephson current 
for various values of the ratio of the
gap amplitude to the tunneling matrix element
$\Delta_0 / t_{\perp}$. Data from the recent cross-whisker measurements on
$\rm Bi_2Sr_2CaCu_2O_{8+\delta}$ by
Takano {\it et al.}\cite{takano2} are also shown as filled circles. 
In Fig.~3(a) the tunneling
element is assumed to be constant. Therefore the angular dependence
is less pronounced than in Fig.~3(b) where an additional
$\cos^2{(2 \phi)}$ modulation of $t_{\perp}$ has
been included. Comparing with the experimental data, both models appear to 
fit the strong angular dependence of the measured current reasonably well with 
$\Delta_0 / t_{\perp} \sim 10$ for $F(\phi )=1$, and $\Delta_0 / t_{\perp} 
\sim 20$ for $F(\phi ) = |f_< f_>|^2$. However, the rapid decay 
of $I_c (T= 0 , \alpha )$ appears to be 
better captured by the model which includes the additional angular modulation
of the tunneling matrix element. In this comparison with the experiment we 
have assumed that the data point at $\alpha = 0.488 \pi$ which violates the 
otherwise monotonic behavior in $I_c (T= 0 , \alpha )$ is of extrinsic 
origin, perhaps 
related to inhomogeneities in the contact surfaces. 

\begin{figure}[h]
\vspace{0.9cm}
\centerline{\psfig{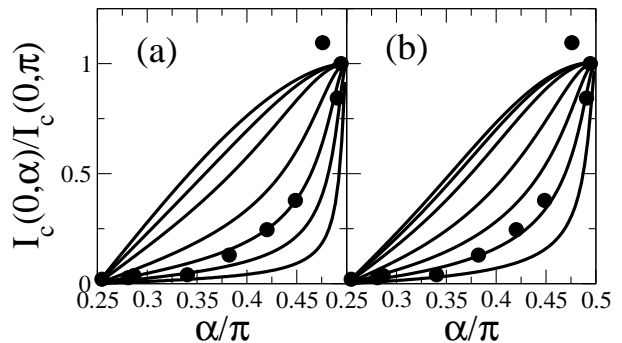}}
\vspace{0.3cm}
\caption{Dependence of the zero-temperature c-axis Josephson current 
on the twist angle $\alpha$. Solid lines: calculated $I_c (T= 0 , \alpha )$
according to Eq.~7, (a) with $F(\phi )=1$, and (b) with 
$F(\phi ) = |f_< f_>|^2$. From top to bottom the ratio of the gap amplitude to 
the c-axis tunneling matrix element is given by $\Delta_0 / t_{\perp}$ = 
0, 1, 2, 5, 10, 20, and 50. Circles: data from Ref. 10.
}
\end{figure}

The most important conclusion from the above considerations is that Andreev
scattering has a crucial role in determining the strong angular dependence 
of the 
Josephson coupling across grain boundaries. As observed in Fig.~3, 
the additional factor $(1 + (\Delta_0 / t_{\perp})^2 (|f_>| - |f_<|)^2)^{-1}$
leads to substantial deviations from the simple proportionality
$I_c (T , \alpha ) \propto \cos{(2
\alpha)}$, which one obtains without taking Andreev processes into account. 
This observation is also related to the 
question of whether these grain boundaries are S-N-S or S-I-S junctions. The 
present study indicates that they seem to be something in between.   

Finally, let us use the above formalism to make some predictions regarding 
the temperature dependence of the c-axis Josephson tunneling current for 
various twist angles. From the discussion of the angular dependence it 
is expected that the current exhibits peaks at $\alpha = 0$ and $\pi /2$,
and vanishes at $\alpha = \pi /4$ and $\alpha = 3\pi /4$. In analogy
to the data presented in Ref. \cite{takano2}, we show in Fig.~4
the temperature dependence for selected twist angles in the region 
$\alpha \in [\pi / 4, \pi/2]$. The ratio $\Delta_0 / t_{\perp}$ was fixed
according to the best fits obtained in Figs.~3(a) and (b). One observes 
in Fig.~4 that while the main temperature dependence is governed by the 
prefactor $\Delta^2(t) \sim (1 - t^3)$ in Eqs.~3 and 6, the most noticeable
differences between the two cases, $F(\phi )=1$ vs. $F(\phi ) = |f_< f_>|^2$,
occur in the low-temperature regions. Moreover, they are more pronounced 
at large twist angles close to $\alpha = \pi /2$. Also, exactly at  
$\alpha = \pi /2$ the expression for the Josephson
current reduces to $\rho_{sc} (t)$.  Comparison of future
experimental data with these curves will aid in distinguishing between 
the proposed models. 

\begin{figure}[h]
\vspace{0.9cm}
\centerline{\psfig{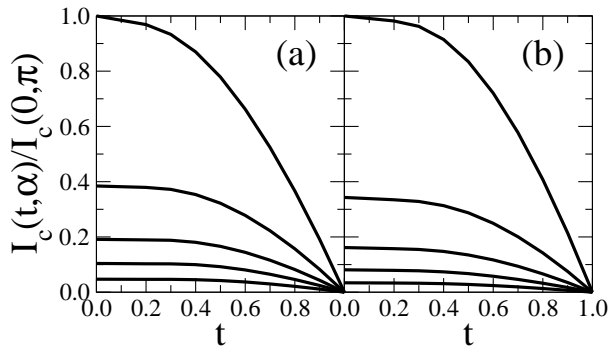}}
\vspace{0.3cm}
\caption{Temperature dependence of the
c-axis Josephson current. From top to bottom the twist angle is 
$\alpha = \pi/2, 0.45\pi , 0.4\pi , 0.35\pi$, and $0.3 \pi$.   
(a) $F(\phi )=1$ and $\Delta / t_{\perp}$ = 10. (b)  
$F(\phi ) = |f_< f_>|^2$ and $\Delta / t_{\perp}$ = 20. 
}
\end{figure}

In conclusion, we have studied the angular and temperature dependence of the 
superfluid density and the Josephson current across c-axis grain boundaries,
such as the recently synthesized twisted cross-whisker junctions of   
$\rm Bi_2Sr_2CaCu_2O_{8+\delta}$.
For both quantities, $\rho_{s c} (t)$ and
$I_c (T , \alpha )$, the inclusion of an additional 
$\cos^2{(2 \phi)}$ angular modulation in the tunneling matrix element 
is important for accurate modeling of the low-temperature properties
of nodal cuprate superconductors without Cu-O chains. 
The observed strong angular dependence of the Josephson current, 
deviating substantially from a simple $\cos{(2\alpha )}$ behavior,
can be accounted for by Andreev scattering processes at the grain boundaries. 
Therefore the measured Josephson current across whisker surfaces in 
$\rm Bi_2Sr_2CaCu_2O_{8+\delta}$ appears to be fully consistent with a 
$\rm d_{x^2-y^2}$-wave symmetry of the superconducting order parameter.

We thank Y. Matsuda, B. Normand, and Y. Takano for useful discussions, 
and acknowledge financial 
support by the National Science Foundation, DMR-0089882.

\end{document}